\documentclass[reprint,amsmath,amssymb,aps,twocolumn]{revtex4-1}

\usepackage{graphicx}% Include figure files
\usepackage{dcolumn}% Align table columns on decimal point
\usepackage{bm}% bold math
\usepackage{subfigure}

\usepackage{xcolor}

\usepackage{booktabs} % 导入三线表需要的宏包
\usepackage{multirow}
\usepackage{makecell}
\usepackage{url}
\usepackage{footnote}
\usepackage{setspace}

\newcommand{\beq}{\begin{equation}}
\newcommand{\eeq}{\end{equation}}
\newcommand{\bq}{\begin{equation}}
\newcommand{\eq}{\end{equation}}
\newcommand{\ba}{\begin{array}}
\newcommand{\ea}{\end{array}}
\newcommand{\beqa}{\begin{eqnarray}}
\newcommand{\eeqa}{\end{eqnarray}}
\newcommand{\beqs}{\begin{subequations}}
\newcommand{\eeqs}{\end{subequations}}

\def\leqq{\leqslant}

\begin{document}
%\pagestyle{plain}
%\setmainfont{Times New Roman}
\preprint{APS/123-QED}

%\setcounter{page}{1}
% \title{Cosmological Constraints on Holographic Dark Energy\\ 
% and Analysis of Cosmic Tensions}
\title{Constraining Holographic Dark Energy and Analyzing Cosmological Tensions}

% Force line breaks with \\

\author{Xin Tang$^{1,2}$}
\author{Yin-Zhe Ma$^{3,4}$}
\email{Corresponding author: Y.-Z. Ma, \url{mayinzhe@sun.ac.za}}
\author{Wei-Ming Dai$^{5}$}
 \author{Hong-Jian He$^{6,7}$}
%\email{ hjhe@sjtu.edu.cn}
 \affiliation{%
$^{1}$Purple Mountain Observatory, CAS, No. 10 Yuanhua Road, Qixia District, Nanjing 210034, China \\
$^{2}$School of Astronomy and Space Science, University of Science and Technology of China, Hefei, Anhui 230026, China \\
$^{3}$Department of Physics, Stellenbosch University, Matieland 7602, South Africa \\
$^{4}$National Institute for Theoretical and Computational Sciences (NITheCS), Stellenbosch, Matieland, 7602, South Africa \\
$^{5}$School of Physical Science and Technology, Ningbo University, Ningbo 315211, China \\
$^{6}$Tsung-Dao Lee Institute \& School of Physics and Astronomy,
Key Laboratory for Particle Astrophysics and Cosmology,
Shanghai Jiao Tong University, Shanghai, China \\
$^{7}$Physics Department \& Institute of Modern Physics, Tsinghua University, Beijing, China
}%

\date{\today}% It is always \today, today,

\begin{abstract}
We investigate cosmological constraints on the holographic dark energy (HDE) using the state-of-the-art cosmological datasets:\ {\it Planck} CMB angular power spectra and weak lensing power spectra, Atacama Cosmology Telescope (ACT) temperature power spectra, baryon acoustic oscillation (BAO) and redshift-space distortion (RSD) measurements from six-degree-field galaxy survey and Sloan Digital Sky Survey (DR12 \& DR16) and the Cepheids-Supernovae measurement from SH0ES team (R22).\ We also examine the HDE model and $\Lambda$CDM with and without $N_{\rm eff}$ (effective number of relativistic species) being treated as a free parameter.\ We find that the HDE model can relieve the tensions of $H_0$ and $S_8$ to certain degrees.\ With ``{\it Planck}+ACT+BAO+RSD'' datasets, the constraints are $H_0 \!=\! 69.70 \pm 1.39\ \mathrm{km\ s^{-1} Mpc^{-1}}$ and $S_8 \!=\! 0.823 \pm 0.011$ in HDE model, which brings down the Hubble tension down to $1.92\sigma$ confidence level (C.L.) and the $S_8$ tension to $(1\!-\!2)\sigma$ C.L. By adding the R22 data, their values are improved as $H_0 = 71.86 \pm 0.93 \,\mathrm{km\ s^{-1} Mpc^{-1}}$ and $S_8 = 0.813 \pm 0.010$, which further brings the Hubble tension down to $0.85\sigma$ C.L. and relieves the $S_{8}$ tension.\ We also quantify the goodness-of-fit of different models with Akaike information criterion (AIC) and Bayesian
information criterion (BIC), and find that the HDE agrees with the observational data better than the $\Lambda$CDM and other extended models (treating $N_{\rm eff}$ as free for fitting).
\end{abstract}

%\keywords{Cosmological parameters; Dark energy}%Use showkeys class option if keyword
\maketitle

%\tableofcontents

\section{Introduction} 
\label{sec:intro}
The standard $\Lambda$ Cold-Dark-Matter ($\Lambda\mathrm{CDM}$) cosmological model has successfully explained the observations of the cosmic microwave background radiation (CMB~\cite{Spergel2003,Spergel2007,Komatsu2009,Komatsu2011,Hinshaw2013,Planck2014_params,
Planck2016_params,plc18-cp}), large-scale distributions of the galaxies~\cite{Cole2005,Beutler2011,Tegmark2004,Alam2021}, and also the relic abundance of the primordial elements formed during the Big-Bang Nucleosynthesis (BBN)~\cite{Cooke2016,Cooke2018}. However, it still has several outstanding problems to be solved. One of the major observational contradictions in recent years is the Hubble tension, which states that the measured $H_{0}$ value from the CMB and the $\Lambda\mathrm{CDM}$ model is lower than the distance-ladder measurements by more than $5\sigma$ confidence level (C.L.)~\cite{Verde2019}. For local measurements, the value of $H_0$ is obtained using distances and redshifts measured from Cepheid in Type Ia supernovae (SN Ia) hosts~\cite{Riess2022}, with no dependence on cosmological models. The recent Cepheids-determined distance ladder from ``SH0ES'' team gives $H_{0}\!=\!73.30\pm 1.04\,{\rm km}\,{\rm s}^{-1}\,{\rm Mpc}^{-1}$~\cite{Riess2022}. In constrast, {\it Planck} CMB experiment~\cite{plc18-cp} and the Atacama Cosmological Telescope (ACT~\cite{act-dr4}) gave $H_0 \!=\! 67.4 \pm 0.5\ \mathrm{km\, s^{-1}\, Mpc^{-1}}$ and $67.9 \pm 1.5\ \mathrm{km\, s^{-1}\, Mpc^{-1}}$ respectively, which is more than $5\sigma$ C.L. smaller than the CMB-measured values.  Another tension is called ``$S_{8}$ tension'', in which $S_{8}=\sigma_{8}\sqrt{\Omega_{\rm m}/0.3}$, where $\sigma_8$ is the {\it rms} (root mean square) density fluctuations on scale of $8\,h^{-1}{\rm Mpc}$, and $\Omega_{\rm m}$ is the fractional matter density. Here $\sigma_8$ and $\Omega_{\rm m}$ are derived parameters which rely on a specific cosmological model, both in CMB and weak lensing measurements. The {\it Planck}+ACT measured $S_{8}$ as $S_8 \!=\! 0.831 \pm 0.023$~\cite{act-dr6}, which is systematically higher than the weak lensing measurements, such as the Dark Energy Survey Year-3 Results (DES-Y3, with
$S_8^{} \!=\! 0.759^{+0.025}_{-0.023}$~\cite{des-y3amon}), Kilo Degree Survey (KiDS-1000, with 
$S_8^{} \!=\! 0.759^{+0.024}_{-0.021}$~\cite{kids-1000}), 
and Subaru Hyper Suprime-Cam Year-3 Results (HSC-Y3, with 
$S_8^{} \!=\! 0.769^{+0.031}_{-0.034}$~\cite{hsc-y3li}).\ 
Hence, the CMB-derived quantity is systematically higher than that measured by using the large-scale-structure datasets.

Different theoretical models beyond $\Lambda$CDM have been proposed to physically explain the $H_{0}^{}$ and $\sigma_{8}^{}$ tensions, but solving the problems ``all in one go'' is not easy because the existing observational data already highly constrain most alternative models.\ For instance, changing the physics prior to the recombination to shrink the acoustic horizon ($r_{\rm d}^{}$) may boost the CMB-measured $H_{0}^{}$ to a higher value (i.e., reducing the Hubble tension), but it may also increase the projected value of $S_{8}$ and make it more incompatible with the large-scale structure measurements.\ This type of solutions includes the primordial magnetic field (PMF~\cite{pmf}) and early dark energy (EDE~\cite{ede}), which have been proven unable to relieve the $S_{8}$ tension~\cite{pmf,Jedamzik2021}. There is also some current research on EDE extension models~\cite{ede2, ede3_Reeves_2023, ede4_Cruz_2023}, showing that they can alleviate the $S_8$ tension by introducing additional conditions or parameters.\ 
Another solution is to have self-interacting neutrinos\,\cite{Cyr-Racine2014, Kreisch2020, He2020}, 
which can delay the onset of neutrino free-streaming until close to the radiation-matter equality and thus accommodates a larger Hubble 
constant\, \footnote{We also notice that Ref.\,\cite{He2020} provided an attractive resolution to the Hubble tension by consistently realizing the neutrino self-interactions, which respects the electroweak gauge symmetry and avoids producing a too large $N_{\rm eff}$ by having  $\Delta N_{\rm eff}\simeq 0.2\,$.}.\ 
A recent such proposal added a fourth ``self-interacting'' neutrino,  but this model requires $N_{\rm eff} \!\!\sim\! 4$ 
which is much higher than the constraints set by {\it Planck} and ACT~\cite{Planck2016_params,act-dr6,Khalife2023}. Other studies that intend to solve the $H_0$ and $S_8$ include, the interacting dark energy model~\cite{Di_Valentino_2020}, the small-scale baryon inhomogeneities (i.e. clumping)~\cite{Rashkovetskyi_2021}, and the mirror twin Higgs model with three extra parameters~\cite{Bansal_2022} etc. For other proposed solutions and theoretical guide, please refer to~\cite{Knox2020,Heisenberg_2022,Cardona_2023}.

%\cite{Heisenberg_2022}, they analysed the late-time extensions that an EoS should have to solve the tensions but didn't propose a solution. Ref.\cite{Cardona_2023} also discussed the DE model based on HP principles, but unlike the parameterised approach to HDE that we used, with more parameters, and with a less up-to-date and richer dataset than ours.}

\vspace*{1mm}

Different from other dark energy (DE) models, the holographic dark energy (HDE) was invoked from the idea that in quantum field theory, the infrared cut-off should exclude all microscopic states lying within the Schwarzschild radius\,\cite{cohen1999}.\ Hence, by relating the ultraviolet (UV) cut-off at a short distance to the infrared (IR) cut-off at a large distance, one can formulate the dark energy with only one extra free parameter ($c$). In this holographic dark energy model, the IR cutoff is taken as the future event horizon of the Universe~\cite{li2004}, and the cosmic coincidence problem can be solved by inflation in this scenario by assuming the minimum number of $e$-foldings. Earlier research confronted the HDE model with the then observational data and showed that it can fit the CMB, Type-Ia supernovae and BAO data slightly better than $\Lambda$CDM~\cite{Huang2004, Ma2009,Ma2010,Li2013}.\ In recent years, the HDE fitting results from Dai, Ma and He (2020)~\cite{dai2020} showed that the HDE model can reconcile the Hubble tension: with  {\it Planck} 2018 CMB data and SDSS-DR12 (Sloan Digital Sky Survey Data Release 12) BAO data, the best-fitting Hubble constant is $H_0 = 71.54 \pm 1.78\,{\rm km}\,{\rm s}^{-1}\,{\rm Mpc}^{-1}$, which is consistent with the local $H_0$ measurement of 2019 LMC Cepheids data (R19)~\cite{Riess2019} at 1.4$\sigma$ confidence limit.

\vspace*{1mm}

In this work, we update the previous constraints by incorporating recent observational data from ACT, SDSS-DR12, SDSS-DR16 (denoted as BAO and RSD), and also the refined determination of $H_{0}$ from 75 Milky Way Cepheids with Hubble Space Telescope (HST) photometry and Gaia EDR3 parallaxes recalibrated distance ladders (denoted as R22~\cite{Riess2022}).\ We will also treat the effective number of relativistic species ($N_{\rm eff}$) as a free parameter in our likelihood analysis and examine whether the fitting results can be improved or not.\ We want to inspect whether the HDE model can play  
the following three roles altogether:
\begin{itemize}
\item[{\Large $\bullet$}] 
to reconcile the Hubble tension between the recent CMB data ({\it Planck} and ACT) and BAO from galaxy surveys (SDSS-DR16), and the distance-ladder measured $H_{0}$ value (R22);

\item[{\Large $\bullet$}] 
to reconcile the $S_{8}$ tension between the CMB and BAO measured values and the large-scale structure measured values (galaxy and CMB lensing);

\item[{\Large $\bullet$}] 
to allow a slightly larger value of $N_{\rm eff}$ that can accommodate the standard model prediction ($N_{\rm eff}=3.046$~\cite{Mangano2005,Salas2016,Akita2020,Froustey2020,Bennett2021}).

\end{itemize}

The rest of the paper is organised as follows. In Sec.\,\ref{hde}, we present the theoretical formulation and predictions of HDE model.\ In Sec.\,\ref{data}, we show the observational data we will use and 
explain the methodology of data fitting.\ In Sec.\,\ref{Result}, we present our results and discuss the implications.\ 
Finally, we conclude in Sec.\,\ref{conclusion}.

\section{Holographic Dark Energy} 
\label{hde}
Holographic dark energy (HDE) was proposed in light of the holographic principle (HP)\,\cite{hooft1993}, which states that all the information contained in a space of with scale $L$ can be described by a theory at the boundary of the space~\cite{susskind1995}.\ This principle is regarded as one of the most important fundamental principles of quantum gravity, and holds the promise to provide the correct description of the quantum gravity at microscopic scales.\ But, the conventional local quantum field theory does not agree with HP when including black hole into the system or at the scale of its Schwarzschild radius because most quantum states would collapse.\ Cohen, Kaplan and Nelson (1999) avoided this catastrophe by suggesting that the energy within a space of the Schwarzschild radius $L$ should not exceed the mass of a black hole of the same size, which means $\,L^3\Lambda^4 \!\leqslant\! LM^{2}_{\rm Pl}\,$~\cite{cohen1999}, where $\Lambda$ represents the UV cutoff and $M_{\rm Pl} = (8\pi G)^{-1/2}$ is the reduced Planck mass.\ Thus, a UV cutoff scale $\Lambda$ is related to an IR cutoff $L$ due to the limit set 
by avoiding the formation of a black hole.\ 
This requires the enclosed vacuum energy to satisfy: 
$\rho_{\rm de}^{} \!\!\sim\! \Lambda^4 \!\leqslant\! 
 M^{2}_{\rm Pl}L^{-2}$.\ 
Taking the entire universe into account and regarding the vacuum energy as dark energy, Li (2004) suggested that the holographic dark energy density is given by saturating the above inequality up to 
a coefficient $c\,$ as free parameter~\cite{li2004}:
\begin{eqnarray}
\rho_{\rm de}^{} = 3c^2M^{2}_{\rm Pl}L^{-2},
\label{eq:rho_de}	
\end{eqnarray}
where $3c^2$ is a numerical factor for the convenience of calculation, as introduced in Li (2004)~\cite{li2004}. $L$ is the future event horizon, which is an assumption chosen to obtain an accelerated expansion of the universe
\begin{eqnarray}
L = R_{\rm eh}^{} = a\int_{t}^{\infty}\! \frac{{\rm d} t'}{\,a(t')\,} 
%\nonumber \\
= a \int_{a}^{\infty}\!\!\frac{{\rm d}a'}{\,H(a')a'^2~}\,,
\label{e2}
\end{eqnarray} 
with $a\!=\!a(t)$ and $a'\!=\!a(t')\,$.

Refs.\,\cite{li2004,Ma2009} further showed that Eq.~(\ref{e2}) leads to the DE equation of the state (EoS) being close to $-1$\,.\ 
Then, from Friedmann equations we have
\beqs 
\begin{eqnarray}
3M^{2}_{\rm pl}H^{2} &=& \rho_{\rm m}^{}+\rho_{\rm de}^{}\,, 
\\[1mm]
\frac{{\rm d}H}{{\rm d}t} &=& 
-\frac{1}{2M^{2}_{\rm Pl}}\left(\rho_{\rm m}^{}+\rho_{\rm de}^{}
+\mathcal{P}_{\rm de}^{} \right) .~~~~
\end{eqnarray}
\eeqs
The energy conservation requires that the covariant derivative acting on the energy-momentum tensor should vanish, 
and this leads to the following conditions,
\beqs 
\begin{eqnarray}
\frac{{\rm d}\rho_{\rm m}}{{\rm d}t}+3H\rho_{\rm m} &=& 0 \,, 
\\
\frac{{\rm d}\rho_{\rm de}}{{\rm d}t}+3H\left(\rho_{\rm de}+\mathcal{P}_{\rm de} \right) &=& 0\,.
\end{eqnarray}
\eeqs 

One can solve this dynamic system completely.\ 
Utilizing $w_{\rm de}\equiv \mathcal{P}_{\rm de}/\rho_{\rm de}$, and $\rho_{\rm de}=3H^{2}M^{2}_{\rm pl}\Omega_{\rm de}$, we can substitute them and obtain a complete set of differential equations for $\Omega_{\rm de}(z)$ and $H(z)$~\cite{Ma2009,Ma2010}:
\beqs 
\label{eq:Omega_H}
\begin{eqnarray}
\hspace*{-3mm}
\frac{{\rm d}\Omega_{\rm de}(z)}{{\rm d}z}+\frac{\Omega_{\rm de}}{1+z}\left[\left(1-\Omega_{\rm de} \right)\left(1+\frac{2\sqrt{\Omega_{\rm de}}}{c} \right) \right] 
\!\!&=&\! 0\,, 
\\[1.5mm]
\hspace*{-3mm}
\frac{{\rm d}H}{{\rm d}z}+\frac{H(z)}{1+z}\left[\frac{\Omega_{\rm de}}{2}\left(1+\frac{2}{c}\sqrt{\Omega_{\rm de}} \right)-\frac{3}{2}\right] \!\!&=&\! 0\,.~~~~ 
\end{eqnarray}
\eeqs 
The EoS parameter of the dark energy can be solved in the following form:
%\pagebreak 
%
\begin{eqnarray}
w_{\rm de}^{} =  -\frac{1}{3} - \frac{2}{3}
\frac{\rho^{1/2}_{\rm de}}{\sqrt{3\,}cM_{\rm Pl}H\,} 
%\nonumber \\
= -\frac{1}{3} - 
\frac{2}{3}\frac{\sqrt{\Omega_{\rm de}^{}(z)}\,}{c}.~~~
\label{e5}
\end{eqnarray}

As can be seen, at the early universe when 
$\Omega_{\rm de}^{} \!\ll\! 1$, we have 
$w_{\rm de}^{} \!\to\! -1/3$ which is greater than $-1\,$.\  
In this case the universe experiences a quintessential acceleration~\cite{Ratra1988,Wetterich1988,Carroll1988}.\ 
But at the late stage when HDE is dominant, 
$\Omega_{\rm de} \!\rightarrow\! 1\,$, 
and thus $w_{\rm de}^{} \!\rightarrow\! -1/3 \!-\! 2/3c\,$.\ 
As long as $c$ is positive, the expansion of the universe would accelerate according to the second Friedmann equation, 
$\Ddot{a}/a \!=\! -4\pi G\rho\,(1 \!+\! 3w)/3$\,, 
which agrees with the current observations.\ 
Hence, we can fit the parameter $c$ and other cosmological parameters with the observational data of the CMB angular power spectra, galaxy BAOs and BBN to find the best-fitting parameters, and then compare the fitting results with $\Lambda$CDM.\ 
As shown in Dai et al. (2020)\,\cite{dai2020}, the then {\it Planck}+BAO12+R19 data~\footnote{BAO12 refers to the SDSS data release 12 (DR12) of the measurements of $D_{\rm M}r_{\rm fid,d}/r_{\rm d}$ 
at $z_{\rm eff}=0.38, 0.51$ and $0.61$~\cite{Alam2017}.\ R19 refers to the Riess et al.\,(2019)~\cite{Riess2019}'s measurement of $H_{0}$ by using Large Magellanic Cloud Cepheid Standards.} gave the best-fitting value as $c = 0.51 \pm 0.02$ and $H_{0}=73.12\pm 1.14\,{\rm km}\,{\rm s}^{-1}{\rm Mpc}^{-1}$, which resolves the Hubble tension completely. And also, we can observe that $0<c<1$ due to the constraints of observations, i.e., $w_{\rm de}<-1$ in the late universe, indicating that it crosses the phantom divide. The physical reason is that, HDE can cause the Universe to have a smaller acceleration earlier on, and a faster acceleration at the later stage, but still keeps the total angular diameter distance to the last-scattering surface unchanged.\ Therefore, this {\it delayed-and-catch up acceleration} can make the model still fit the CMB angular power spectrum, but enhance the local expansion rate (local $H_{0}^{}$) to match the R19 result as in Dai et al. (2020)~\cite{dai2020}.\ In the following sections, we will analyze the model by using the most up-to-date datasets and present the fitting results.

%This change in the EoS parameter of HDE provides a favorable dynamic characteristic for the $H_0$ tension alleviation, as discussed in the last paragraph of Section 2 by Ref. . And this is the main reason why we chose the HDE model to study cosmic tensions.

%which shows that the future event horizon as the characteristic length scale $L$ is indeed a suitable choice. By selecting the appropriate value of $c$ obtained by observation constraints, we can get the change of $w_{\rm de}$ of HDE with redshift $z$. As we saw in our discussion of the results in Section \ref{Result}, the value of c is between 0 and 1, thus in the later universe, $w$ is less than -1. 

%As stated in the Ref. \cite{li2009}, it must be emphasized that this expression for $\rho_{\rm de}$ is obtained by combining HP and dimensional analysis, unlike other common dark energy models. Due to this unique feature, and the fact that the HDE model was the first DE theoretical model inspired by HP, the HDE model is significantly different from any other DE theory. It also coincides well with current cosmological observations, which makes HDE very competitive.

\section{Data and Methodology} 
\label{data}

\begin{figure*}[htb]
\centering
%\subfigure{
\includegraphics[width=8.4cm]{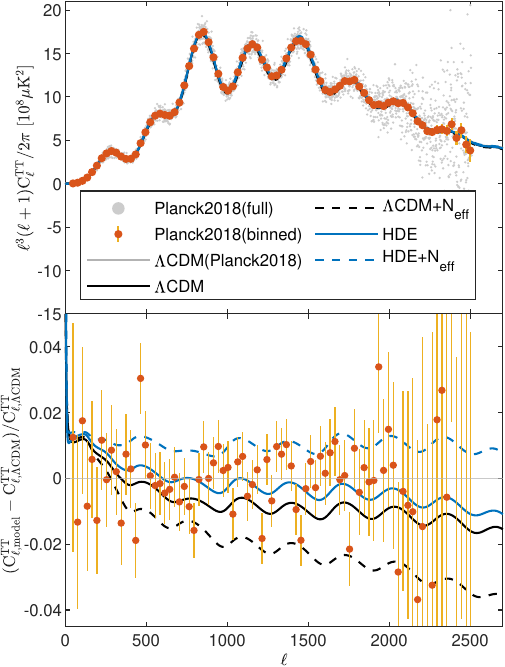}
%\caption{fig1}
%}
%\quad
%\subfigure{
\includegraphics[width=8.4cm]{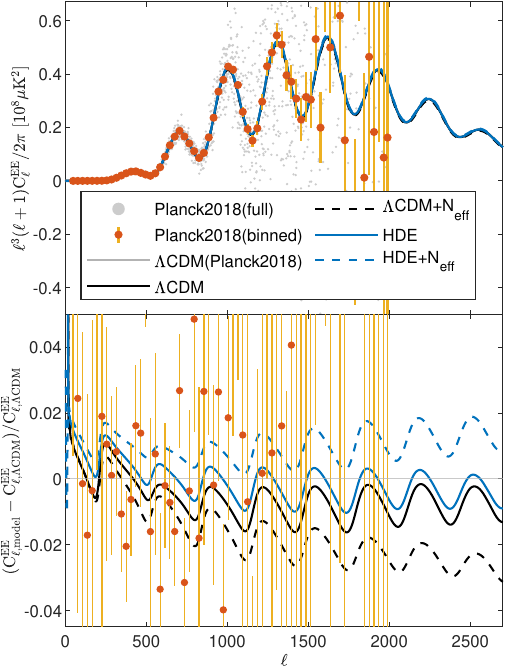}
%}
\vspace*{-2mm}
\caption{CMB TT (left) and EE (right) power spectrum with parameters constraints by \textit{Planck}+ACT+BAO+RSD+R22 data combination, which illustrates the effects of $\Lambda$CDM, $\Lambda$CDM+$N_{\mathrm{eff}}$, HDE, and HDE+$N_{\mathrm{eff}}$ on the phase and amplitude of the power spectra.\ Measurements from the {\it Planck} 2018 data release are included, and the grey solid line is the best-fitting $\Lambda$CDM CMB power spectra from the baseline {\it Planck} TT, TE, EE+lowE+lensing ($2 \leqq \ell \leqq 2508$).}
\label{fig:CMBCell}
\end{figure*}

We modify the Boltzmann code {\sc camb}\,\cite{camb} to embed the HDE model [Eq.\,(\ref{eq:Omega_H})] as the background expansion model to compute the angular power spectra of CMB (see also Fig.\,\ref{fig:CMBCell}). We have only considered the effect of the HDE on the background expansion, i.e., we have replaced the components of DE ($\rho_{\rm de}$) in the background density ($\rho$) of the perturbation equations as well as its EoS $w_{\rm de}$, without considering its perturbation term ($\delta_{\rm de} = 0$.\footnote{Notice that perturbation of HDE itself and its interactions are considered in~\cite{Wang_2017}, and Chunshan Lin (2021)~\cite{Lin_2021} introduced a covariant local field theory for the HDE model, providing a method to study its perturbations.}

We then utilize the Markov Chain Monte Carlo (MCMC) technique from the public code {\sc CosmoMC}~\cite{cosmomc} to constrain the parameters of models by using the latest {\it Planck} and ACT CMB data, SDSS DR16 BAO and RSD data and also the R22 data.\ The sampling method we used in {\sc CosmoMC} is a modified Metropolis algorithm, specifically a fast-slow dragging algorithm\cite{neal2005takingbiggermetropolissteps, Lewis_2013}. This approach is suitable for likelihood with numerous fast nuisance parameters like {\it Planck}.
The benchmark models we fit are $\Lambda$CDM and HDE.\ We also add $N_{\mathrm{eff}}^{}$ (the effective number of relativistic species) 
as a free parameter to the fit of each model.\ 
So we have four models ($\Lambda$CDM, HDE, $\Lambda$CDM+$N_{\rm eff}$, and HDE+$N_{\rm eff}$) for the present analysis.\ 
The $\Lambda$CDM model has six cosmological parameters $\Omega_{\rm b}h^2$, $\Omega_{\rm c}h^2$, $100\theta_{\mathrm{MC}}$, $\tau$, $\ln (10^{10}A_{\rm s})$, and $n_{\rm s}^{}$, which correspond to the fractional baryon and dark matter densities, the angular size to the last-scattering surface, the optical depth, the amplitude, and the tilt of primordial curvature perturbations, respectively.\ The HDE model has one more free parameter $c$ as shown in the Eq.\,(\ref{eq:rho_de}).

\begin{table*}[]
\caption{Measurements used in the two datasets BAO and BAO+RSD 
for the present analysis.\ The 6dF $D_{\rm V}$ measurement 
is from Beutler et al.\,(2011)\,\cite{6df} 
and the other measurements are all from Table\,III of Alam et al.\,(2021)\,\cite{Alam2021}.}
\vspace*{2mm}
\begin{ruledtabular}
\begin{tabular}{cccccc}
     & 6dF & MGS & BOSS galaxy (DR12) & BOSS galaxy (DR12) & eBOSS LRG (DR16)\\
    $z_{\mathrm{eff}}$ & 0.106 & 0.15 & 0.38 & 0.51 & 0.698\\ \hline
    \multicolumn{6}{c}{\multirow{2}{*}{\textbf{Dateset:} BAO}}\\
    ~\\
    $r_{\rm d}/D_{\rm V}(z_{\rm{eff}})$ & $0.336 \pm 0.015$ & & & &\\
    $D_{\rm V}(z_{\rm{eff}})/r_{\rm d}$ &  & $4.47 \pm 0.17 $ & & &\\
    $D_{\rm M}(z_{\rm{eff}})/r_{\rm d}$ &  &  & $10.23 \pm 0.17$ & $13.36 \pm 0.21$ & $17.86 \pm 0.33$ \\
    $D_H(z_{\rm{eff}})/r_{\rm d}$ &  &  & $25.00 \pm 0.76$ & $22.33 \pm 0.58$ & $19.33 \pm 0.53$ \\ \hline
    \multicolumn{6}{c}{\multirow{2}{*}{\textbf{Dateset:} BAO+RSD}}\\
    ~\\
    $r_{\rm d}/D_{\rm V}(z_{\rm{eff}})$ & $0.336 \pm 0.015$ & & & &\\
    $D_{\rm V}(z_{\rm{eff}})/r_{\rm d}$ &  & $4.47 \pm 0.17$ & & &\\
    $D_{\rm M}(z_{\rm{eff}})/r_{\rm d}$ &  &  & $10.27 \pm 0.15$ & $13.38 \pm 0.18$ & $17.65 \pm 0.30$ \\
    $D_H(z_{\rm{eff}})/r_{\rm d}$ &  &  & $24.89 \pm 0.58$ & $22.43 \pm 0.48$ & $19.78 \pm 0.46$ \\
    $f\sigma_8(z_{\rm{eff}})$ &  &  & $0.497 \pm 0.045$ & $0.459 \pm 0.038$ & $0.473 \pm 0.041$ \\
\end{tabular}
\end{ruledtabular}
\label{tab:dataset}
\end{table*}

\begin{figure*}[t]
%\centering
\hspace*{-6mm}
\includegraphics[width=17.5cm]{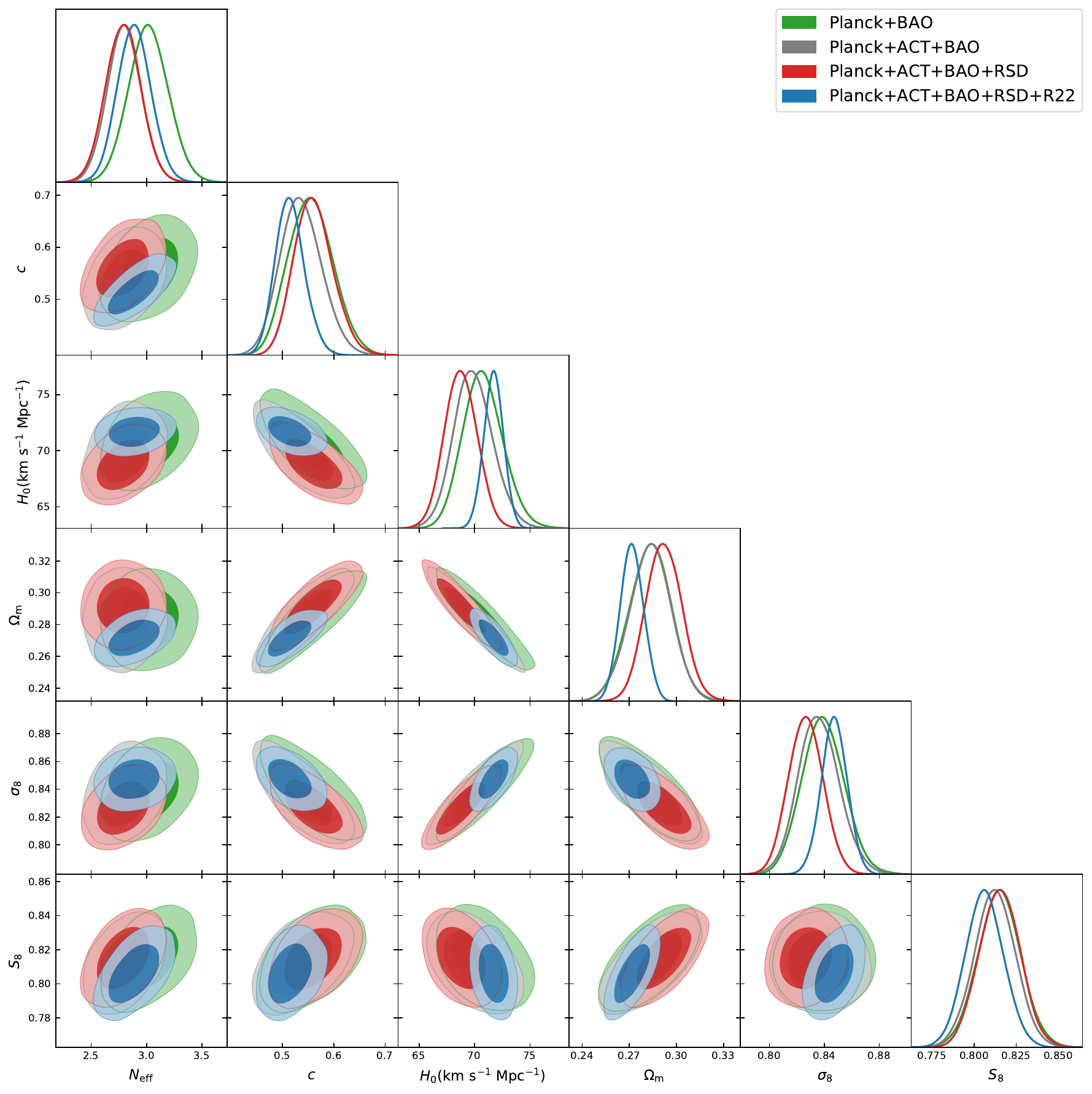}
\caption{Posterior distributions of HDE+$N_{\mathrm{eff}}$ model for selection parameters for four of the dataset combinations, 
at 68\% C.L.\ ($1\sigma$) and 95\% C.L. ($2\sigma$).\ The first two parameters ($N_{\rm eff}$ and $c$) are from the model's base parameter set (Table\,\ref{tab:parameters}), and the last four parameters ($H_{0}^{}$, $\Omega_{\rm m}$, $\sigma_8$ and $S_{8}^{}$) are the derived parameters.}
\label{fig:HDENeff_contours}
\end{figure*}

\begin{figure*}[!htb]
\centering
\subfigure[\,Data set: \textit{Planck}+ACT+BAO+RSD]{
\includegraphics[scale=0.55]{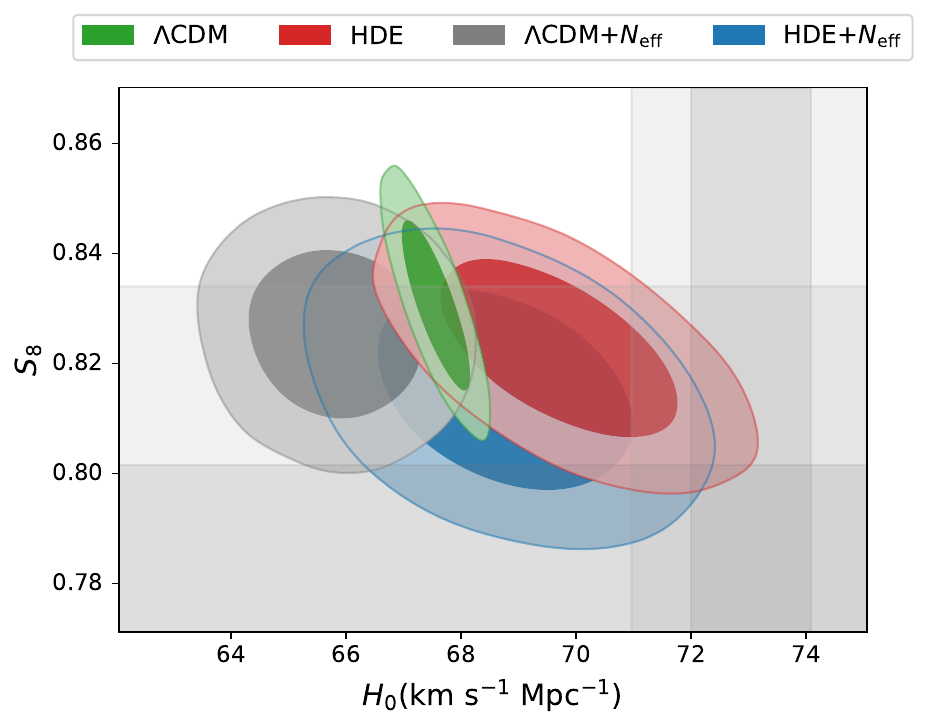}}
%\caption{fig1}
%}
%\quad
\subfigure[\,Data set: \textit{Planck}+ACT+BAO+RSD+R22]{
\includegraphics[scale=0.55]{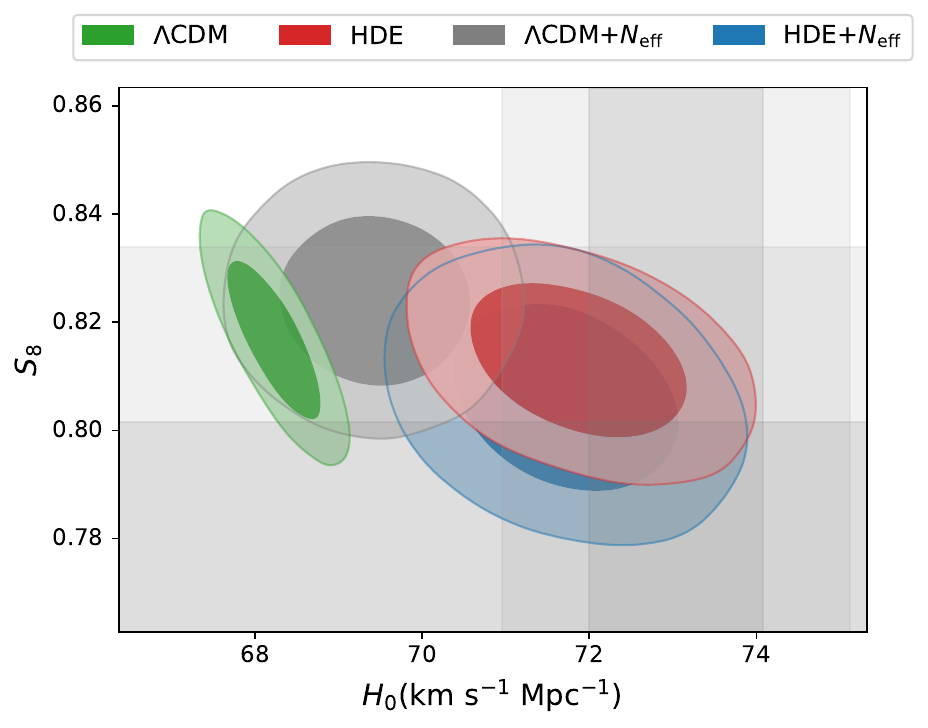}}
\hspace*{-8mm}

\vspace*{-3mm}
\caption{2D marginalized posterior contours in the plane of 
$H_0^{}$ and $S_8^{}$ (68\%\,C.L. and 95\%\,C.L.), 
by using two data combinations ``\textit{Planck}+ACT+BAO+RSD'' [panel-(a)] and ``\textit{Planck}+ACT+BAO+RSD+R22'' [panel-(b)] in the four models $\Lambda$CDM, $\Lambda$CDM+$N_{\mathrm{eff}}$, HDE, and HDE+$N_{\mathrm{eff}}$.\ The gray horizontal and vertical bands 
are the $1\sigma$ and $2\sigma$ regions of the measurements 
$S_8^{} = 0.769^{+0.031}_{-0.034}$ (at $1\sigma$ level)
\cite{hsc-y3li} and $H_0^{} = 73.04 \pm 1.04\,\mathrm{km\ s^{-1} Mpc^{-1}}$ (at $1\sigma$ level) \cite{Riess2022}.}
\label{fig:H0_S8}
\end{figure*}

\begin{table*}[t]
%  \centering
\caption{%
Constraints of 
\textit{Planck}+ACT+BAO+RSD and \textit{Planck}+ACT+BAO+RSD+R22
datasets on the $\Lambda$CDM, $\Lambda$CDM+$N_{\rm eff}^{}$, HDE, and HDE+$N_{\rm eff}^{}$ models.\  For each dataset, the upper block is for the constraints on the fundamental parameters, and the lower block is for the derived parameters.\ The quoted error is given at the $68\%$\,C.L.}
\vspace*{2mm}
\begin{ruledtabular}
\begin{tabular}{lllll}%其中，tabular是表格内容的环境；c表示centering，即文本格式居中；c的个数代表列的个数
    %\toprule[2pt]%设置线宽 
   Parameters & $\Lambda$CDM & $\Lambda$CDM+$N_{\mathrm{eff}}$ & HDE & HDE+$N_{\mathrm{eff}}$\\
\hline     
     %\toprule %[2pt]设置线宽    
%     \midrule %[2pt]
\multicolumn{5}{c}{\multirow{2}{*}{\textbf{Dateset:} {\it Planck}+ACT+BAO+RSD}}\\
~\\
    $\Omega_{\rm b}h^2$ & $0.02236\pm0.00012 $& $0.02215\pm0.00017 $ & $0.02242\pm0.00012 $& $0.02220\pm0.00019 $\\ %\hline 
    
    $\Omega_{\rm c}h^2$ & $0.11962\pm0.00090$ & $0.11545\pm0.00238$ & $0.11876\pm0.00098$& $0.11521\pm0.00248$\\ %\hline    
    
    $100\,\theta_{\rm MC}$  & $1.04111\pm0.00028$ & $1.04159\pm0.00038$& $1.04122\pm0.00027$ & $1.04163\pm0.00039$\\ %\hline   
    
    $\tau$  & $0.0543\pm0.0070$ & $0.0537\pm0.0070$ & $0.0569\pm0.0076$& $0.0555\pm0.0074$\\ %\hline    
    
    $\ln(10^{10}A_{\rm s})$  & $3.051\pm0.013$& $3.037\pm0.015$ & $3.055\pm0.014$ & $3.041\pm0.017$\\ %\hline    
    
    $n_{\rm s}$ & $0.9684\pm0.0034$ & $0.9581\pm0.0064$ & $0.9700\pm0.0036$& $0.9601\pm0.0074$\\ %\hline    
    
    $c$ & - & - & $0.58484\pm0.04158$& $0.56174\pm0.04594$\\ %\hline
    
     $N_{\mathrm{eff}} $ &  $3.046$\,(fixed)  & $2.78\pm 0.14$ & $3.046$\,(fixed) & $2.80\pm 0.16$\\ %\hline    
     \hline
%     \hline[dashed]
%    \cdashline{1-5}   
    $H_0[\mathrm{km\,s^{-1}\,Mpc^{-1}}]$ & $67.55 \pm 0.39 $ & $65.84 \pm 0.99 $ & $69.70 \pm 1.39 $ & $68.78 \pm 1.52 $ \\ %\hline   
    
    $\Omega_{\rm de}$ & $0.6874\pm0.0054$ & $0.6810\pm0.0064$ & $0.7077\pm0.0118$ & $0.7079\pm0.0122$ \\ %\hline    
    
    $\Omega_{\rm m}$ & $0.3126\pm0.0054$ & $0.3190\pm0.0064$ & $0.2923\pm0.0118$ & $0.2921\pm0.0122$ \\ %\hline    
    
    $\sigma_8$ & $0.8139\pm0.0056$ & $0.8007\pm0.0089$ & $0.8337\pm0.0119$ & $0.8269\pm0.0129$\\ %\hline    
    
    $S_8\equiv \sigma_{8}\sqrt{\Omega_{\rm m}/0.3}$ & $0.831\pm0.010$ & $0.826\pm0.010$ & $0.823\pm0.011$ & $0.816\pm0.012$\\ \hline  
    \multicolumn{5}{c}{\multirow{2}{*}{\textbf{Dateset:} {\it Planck}+ACT+BAO+RSD+R22}}\\
~\\
$\Omega_{\rm b}h^2$ & $0.02249\pm0.00012 $& $0.02264\pm0.00014 $ & $0.02241\pm0.00012 $& $0.02227\pm0.00019 $\\ %\hline 
    
    $\Omega_{\rm c}h^2$ & $0.11820\pm0.00084$ & $0.12179\pm0.00217$ & $0.11900\pm0.00097$& $0.11684\pm0.00237$\\ %\hline    
    
    $100\,\theta_{\rm MC}$  & $1.04131\pm0.00027$ & $1.04092\pm0.00034$& $1.04119\pm0.00028$ & $1.04143\pm0.00038$\\ %\hline   
    
    $\tau$  & $0.0587\pm0.0072$ & $0.0580\pm0.0072$ & $0.0553\pm0.0072$& $0.0538\pm0.0073$\\ %\hline    
    
    $\ln(10^{10}A_{\rm s})$  & $3.056\pm0.014$& $3.065\pm0.015$ & $3.052\pm0.014$ & $3.042\pm0.017$\\ %\hline    
    
    $n_{\rm s}$ & $0.9716\pm0.0033$ & $0.9786\pm0.0051$ & $0.9696\pm0.0035$& $0.9634\pm0.0072$\\ %\hline    
    
    $c$ & - & - & $0.53625\pm0.02700$& $0.51616\pm0.03197$\\ %\hline
    
     $N_{\mathrm{eff}} $ &  $3.046$\,(fixed)  & $3.26\pm 0.12$ & $3.046$\,(fixed) & $2.89\pm 0.15$\\ %\hline    
     \hline
    $H_0[\mathrm{km\,s^{-1}\,Mpc^{-1}}]$ & $68.23 \pm 0.37 $ & $69.41 \pm 0.75 $ & $71.86 \pm 0.93 $ & $71.70 \pm 0.93 $ \\ %\hline   
    
    $\Omega_{\rm de}$ & $0.6963\pm0.0049$ & $0.6989\pm0.0050$ & $0.7248\pm0.0076$ & $0.7280\pm0.0080$ \\ %\hline    
    
    $\Omega_{\rm m}$ & $0.3037\pm0.0049$ & $0.3011\pm0.0050$ & $0.2752\pm0.0076$ & $0.2720\pm0.0080$ \\ %\hline    
    
    $\sigma_8$ & $0.8119\pm0.0058$ & $0.8225\pm0.0083$ & $0.8489\pm0.0096$ & $0.8473\pm0.0097$\\ %\hline    
    
    $S_8\equiv \sigma_{8}\sqrt{\Omega_{\rm m}/0.3}$ & $0.817\pm0.010$ & $0.824\pm0.010$ & $0.813\pm0.010$ & $0.807\pm0.011$\\ %\hline 
    %\hline[2pt]
    %\bottomrule[2pt]     
    \end{tabular}
    \end{ruledtabular}
    \label{tab:parameters}
\end{table*}

\begin{figure}[!htb]
    \centering
    \includegraphics[scale=0.45]{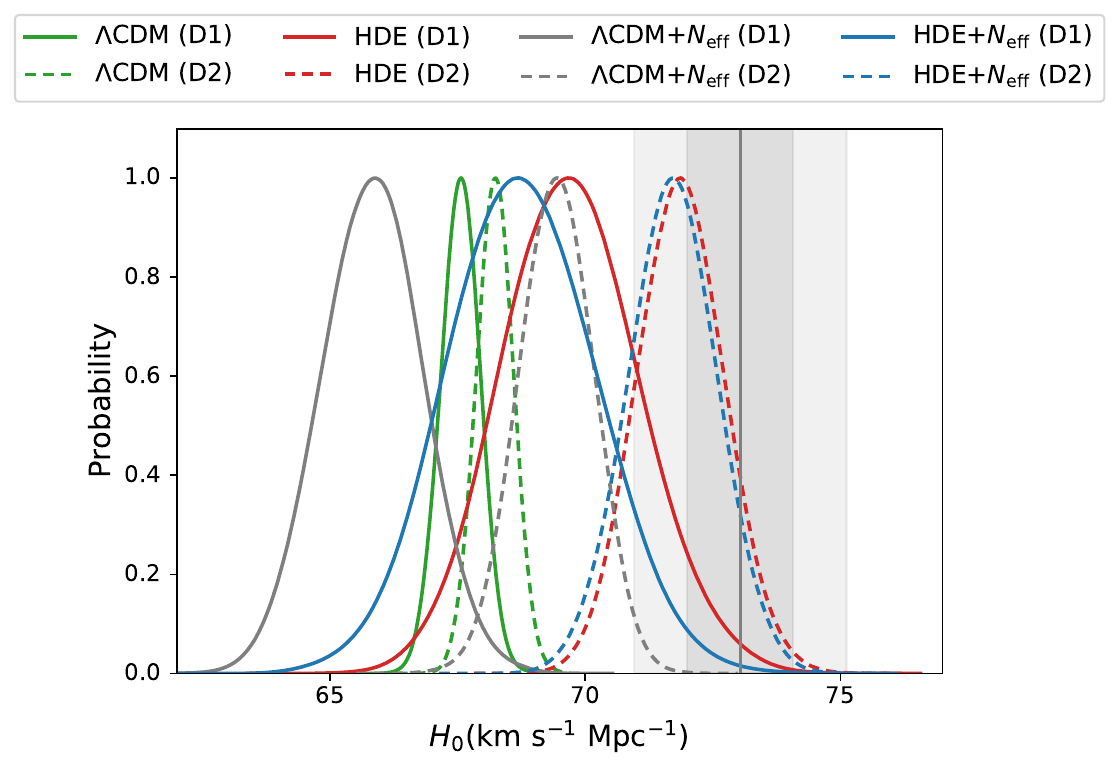}
    \caption{Marginalized posterior distributions for $H_0$ using two data set constraints D1 and D2, where D1 is \textit{Planck}+ACT+BAO+RSD (solid curves) and D2 is \textit{Planck}+ACT+BAO+RSD+R22 (dashed curves). The vertical grey line is the R22 measurements of $H_0= 73.04 \pm 1.04\ \mathrm{km\ s^{-1}\ Mpc^{-1}}$, and the bands response its regions allowed at 1$\sigma$ (dark grey) and 2$\sigma$ (light grey) C.L.}
    \label{fig:H0compare}
\end{figure}

\begin{figure}[!htb]
\centering
\includegraphics[scale=0.45]{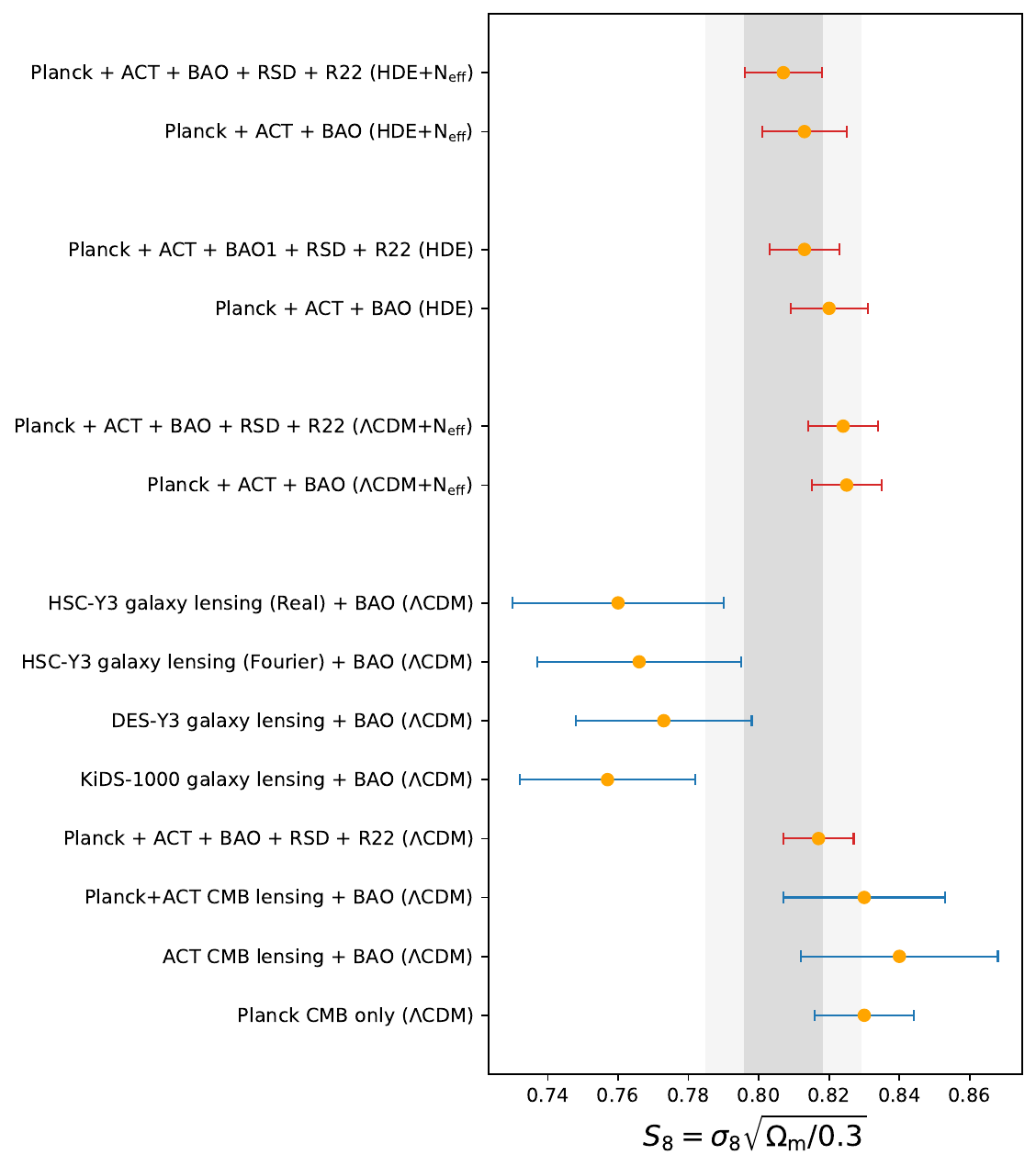}
\vspace*{-3mm}
\caption{Marginalized constraints on $S_8$ ($68\%$\,C.L.) of different weak lensing, CMB and BAO measurements, and their combinations.\ Here we compare our results (red) with the constraints (blue) taken from Table~2 of ACT DR6 lensing\,\cite{act-dr6}.\ In this plot, {\it Planck} CMB only refers to the {\it Planck} data mentioned in Sec.\,\ref{data} above without lensing, ACT CMB lensing refers to the ACT lensing measurement from ACT DR6 lensing, and {\it Planck}+ACT CMB lensing refers to 
lensing-alone data from {\it Planck} and ACT DR6.\ 
The gray bands are the $1\sigma$ and $2\sigma$ confidence limits 
on the model HDE+$N_{\mathrm{eff}}^{}$ by using the dataset \textit{Planck}+ACT+BAO+RSD+R22, (i.e., $S_{8}=0.807 \pm 0.011$ ($1\sigma$ C.L.) in the bottom line, right column of Table\,\ref{tab:parameters}).}
\label{fig:S8constraints}
\end{figure}

\vspace*{1mm}

Then, we explain each dataset used in our fitting:
\begin{itemize}

\item[{\Large $\bullet$}] 
{\boldmath{${\it Planck}$}}: 
We use the CMB high-$\ell$ TT, TE and EE {\tt plik} likelihood, low-$\ell$ temperature {\tt Commander} likelihood, low-$\ell$ {\tt SimAll} EE polarization from the {\it Planck} 2018 data release~\cite{plc18-lk}, and {\it Planck} 2018 CMB lensing power spectrum~\cite{plc18-ls}.\ In the following, we use \textit{Planck} to denote the combination of all {\it Planck} data mentioned here.

\item[{\Large $\bullet$}]  
\textbf{ACT}: We use the likelihood computed from the CMB power spectrum on scales of $\,350\!<\! \ell\!<\! 4000$ 
measured by the ACTPol instrument of the Atacama Cosmology Telescope, derived from seasons 2013-2016 of temperature and polarization data release 4 (DR4)\,\cite{act-dr4}. 
%These spectra have already been marginalized over SZ and foreground emission, and calibration and beam uncertainties. An overall polarization efficiency is the only nuisance parameter required to be sampled.

\item[{\Large $\bullet$}] 
\textbf{BAO}: We use the Baryon Acoustic Oscillation measurements at different redshifts as the standard ruler to constrain cosmological parameters. The Six-degree Field Galaxy Survey (6dF) measures $D_{\rm V}$ at $z_{\mathrm{eff}} = 0.106$~\cite{6df} and the ``Main Galaxy Sample'' (MGS) from Sloan Digital Sky Survey Data Release 12 (SDSS-DR12) measures $D_{\rm V}$ at $z_{\mathrm{eff}} = 0.15$~\cite{mgs}. We also add $D_{\rm M}(z_{\mathrm{eff}})/r_{\rm d}$ and $D_H(z_{\mathrm{eff}})/r_{\rm d}$ measurements at 
$z_{\mathrm{eff}} = 0.38$ and $0.51$ from SDSS-DR12 (BOSS CMASS galaxies)~\cite{Alam2017} and same quantities at $z_{\mathrm{eff}} = 0.698$ from SDSS-IV DR16 eBOSS Luminous Red Galaxies (LRG)~\cite{Alam2021}. The five samples at corresponding redshifts are shown Table~\ref{tab:dataset}.\ The two SDSS-DR12 samples spanning 
$\,0.2 \!<\! z \!<\! 0.6\,$ 
are correlated due to overlapping in $z$, independent of the DR16 results occupying $\,0.6 \!<\! z \!<\! 1.0\,$.\ 
Hence, we include the full covariance matrix of $D_{\rm M}(z_{\mathrm{eff}})/r_{\rm d}$ and $D_H(z_{\mathrm{eff}})/r_{\rm d}$ in our {\sc CosmoMC} likelihoods (see also Table III in~\cite{Alam2021}).\ We use ``BAO'' to denote the BAO-only dataset at these five effective redshifts (Table\,\ref{tab:dataset}).

The $r_{\rm d}=147.8\,{\rm Mpc}$ mentioned above is the sound horizon at the drag epoch, where the value is set by the fiducial cosmological parameters in the spatially-flat $\Lambda$CDM model\,\footnote{The detailed values are listed in the first row of Table~I in Ref.~\cite{bao16-plus}}.\ $D_{\rm M}(z)$ is the co-moving angular diameter distance, and $D_H(z) = c/H(z)$ is the Hubble radius.\ 
$D_{\rm V}(z)$ is the spherically averaged distance defined as $D_{\rm V}(z) = \big(zD_H(z)D^2_{\rm M}(z)\big)^{1/3}$, where the powers of $2/3$ and $1/3$ approximately account for two transverse and one radial dimension and the extra factor of $z$ is a conventional normalization~\cite{Alam2021}.

\item[{\Large $\bullet$}] 
\textbf{BAO+RSD}: We also adopt the BAO measurements with additional measurement of $f\sigma_{8}(z)$ from redshift-space-distortion (RSD). The RSD arises from the peculiar velocities of distant galaxies due to the inhomogeneous distribution of matters. $f\sigma_8(z)$ is the amplitude of the velocity power spectrum, where $\sigma_8(z)$ is the amplitude of linear matter fluctuations on the comoving scale of $8\,h^{-1}{\rm Mpc}$, and $f(z)\equiv \partial \ln D/ \partial \ln a$ is the linear growth rate of structure ($a = 1/(1+z)$ is the scale factor). The RSD measurement provides direct measurement on $f\sigma_8(z)$~\cite{Kaiser1987,Alam2017,Alam2021}.

Hence, we use the joint likelihood of $D_{\rm M}(z_{\mathrm{eff}})/r_{\rm d}$, $D_H(z_{\mathrm{eff}})/r_{\rm d}$, and $f\sigma_8(z_{\mathrm{eff}})$ at 
$z_{\mathrm{eff}}^{}\!=\! 0.38$, $0.51$ and $0.698$ as determined from the combined BAO and RSD likelihoods~\cite{bao16-plus}, and also the two 6dF and MGS samples (Table\,\ref{tab:dataset}).\ 
The measurements at 
$z_{\mathrm{eff}}^{} \!=\! 0.38$ and $0.51$ are taken from the data presented in Table\,8 of SDSS-DR12\,\cite{Alam2017}, and the measurement at $z_{\rm eff}^{} \!=\! 0.698$ is taken from 
SDSS-DR16\,\cite{Alam2021}.\ Our {\sc CosmoMC} likelihoods include the full covariance matrix of $D_{\rm M}(z_{\mathrm{eff}})/r_{\rm d}$, $D_H(z_{\mathrm{eff}})/r_{\rm d}$ and $f\sigma_8(z)$, as $f\sigma_8(z)$ correlates with the BAO measurement at the same redshift but the two BAO SDSS-DR12 results correlate in between redshifts. We use BAO+RSD to denote the $D_{\rm M}(z_{\mathrm{eff}})/r_{\rm d}$, $D_H(z_{\mathrm{eff}})/r_{\rm d}$, and $f\sigma_8(z_{\mathrm{eff}})$ measurements combined with the two $D_{\rm V}$ measurements from 6dF and MGS/SDSS-DR7 (five samples in total), which are presented at Table\,\ref{tab:dataset}.

\item[{\Large $\bullet$}] 
\textbf{R22}: We use the measurement of Hubble constant from SH0ES team as $H_0^{}\!=\! 73.04 \pm 1.04$\,km $\mathrm{s^{-1} Mpc^{-1}}$ by Riess \textit{et al.}\,\cite{Riess2022}.\ This value is the baseline result from the Cepheid-SN Ia sample including systematic uncertainties and lies near the median of all analysis variants.

\end{itemize}

To compare the effects of different datasets on the cosmological parameter constraints, we use the following dataset combinations: ``\textit{Planck}+BAO'', ``\textit{Planck}+ACT+BAO'', ``\textit{Planck}+ACT+BAO+RSD'', ``\textit{Planck}+ACT+BAO+RSD+R22''.\  We also compare the results of the constraints for four different models $\Lambda$CDM, HDE, $\Lambda$CDM+$N_{\mathrm{eff}}^{}$, and HDE+$N_{\mathrm{eff}}^{}$ (the latter two treat $N_{\rm eff}^{}$ as free parameter in the fits), and use the two datasets ``\textit{Planck}+ACT+BAO+RSD'' and ``\textit{Planck}+ACT+BAO+RSD+R22'' for the model comparison.

In this work, we do not use the Pantheon Type-Ia supernovae samples which comprise $1048$ data spanning the redshift range 
$0.01 \!\!<\!\! z \!<\! 2.3$ \cite{pantheon_2018}.\ 
The reason is the possible inconsistency of the parameter constraints due to the {\it large correlation} between high-$z$ ($z\!>\!0.2$) and low-$z$ ($z \!<\! 0.2$) samples of Pantheon subsamples.\ 
As a result, the joint constraints on HDE's $w_{\rm de}(z\!=\!0)$ 
and $H_0^{}$ from individual subsamples and the full Pantheon sample become inconsistent with each other (as shown by Fig.\,3 of Dai et al. (2020)\,\cite{dai2020}).\ In Ref.\,\cite{Valentino2021}, it shows that ``{\it Planck}+BAO'' and ``{\it Planck}+Pantheon'' give inconsistent results by more than 95\% C.L., also suggesting that there might be some uncounted systematics in the Pantheon dataset.

Currently, the new Pantheon+ data has been released, including 1701 light curves of 1550 distinct Ia supernovae covering a redshift range from $z = 0.001$ to $2.26$\cite{Scolnic_2022}, which represents a greater number of Ia SN measurements compared to Pantheon. We present the findings of constraints resulting in the Appendix from a combination of Pantheon+ data with {\it Planck}, ACT, and BAO+RSD utilized in our work.

\section{Results and Discussions} 
\label{Result}

\begin{figure}[ht]
\centering
\includegraphics[width=9cm]{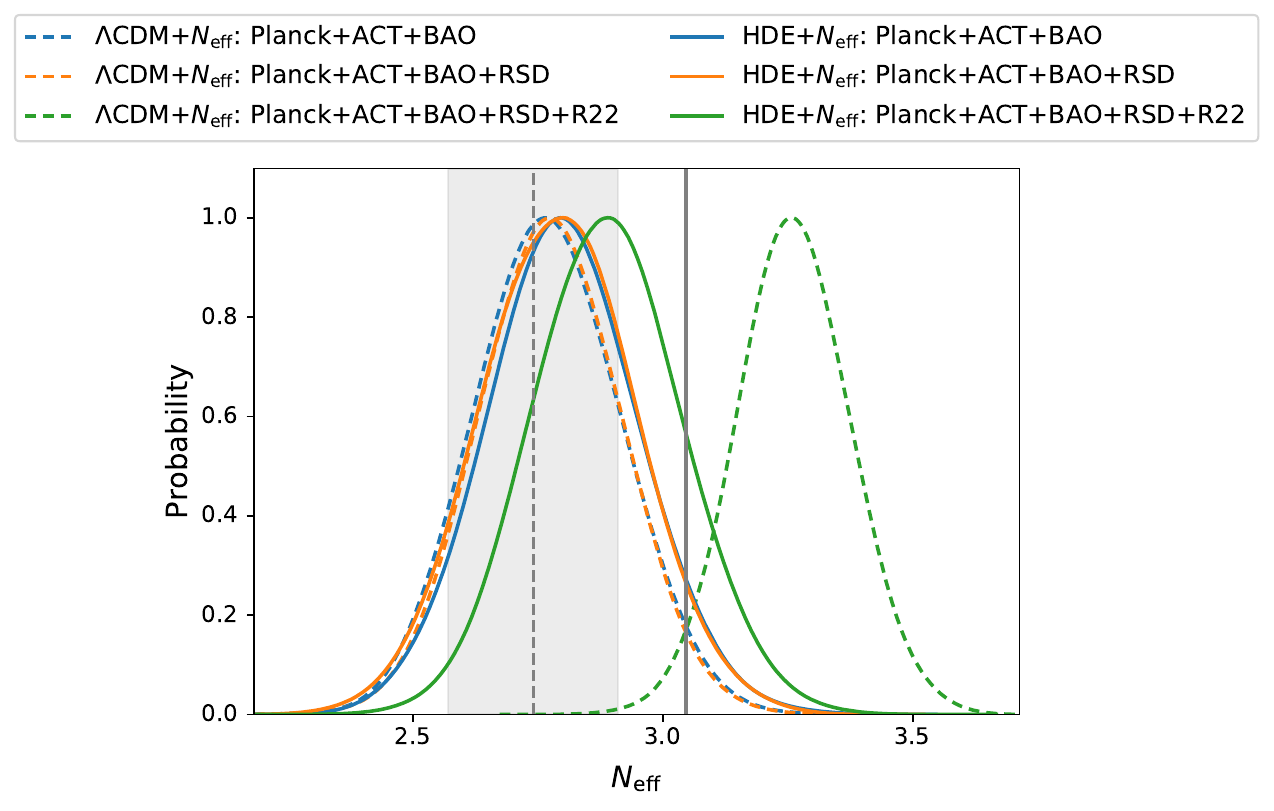}
\vspace*{-6mm}
\caption{%
Constraints on the parameter $N_{\rm eff}^{}$ in $\Lambda$CDM+$N_{\mathrm{eff}}$ model (dashed curves) and HDE+$N_{\mathrm{eff}}$ model (solid curves) 
from the three combined data sets.\  
The gray vertical dashed line with the band is the constraint from {\it Planck}+ACT data on $\Lambda$CDM+$N_{\rm eff}^{}$ model (ACT DR4\,\cite{act-dr4}): $N_{\mathrm{eff}}^{} = 2.74 \pm 1.3\,$.\ 
The gray solid line indicates $N_{\mathrm{eff}}^{} = 3.046$ 
as computed in the standard model of particle physics\,\cite{Mangano2005,Salas2016,Akita2020,Froustey2020,Bennett2021}.}
\label{fig:Neff}
\end{figure}

\begin{table}
%    \centering
\caption{Flat priors of cosmological parameters in MCMC.}
\vspace*{3mm}
%    \setlength{\tabcolsep}{6pt}
%    \small
%    \renewcommand{\arraystretch}{1.2}
%    \resizebox{8.6cm}{!}{
%    \begin{ruledtabular}
    \begin{tabular}{|l|l|}
%    \toprule[2pt]
    \hline
    Parameters & Prior range\\ \hline
    \toprule %[2pt]
       
     $c$ & $[0.2, 2]$\\ %\hline
    
     $\Omega_{\rm b}h^2$ & $[0.005, 0.1]$\\ %\hline
    
     $\Omega_{\rm c}h^2$ & $[0.001, 0.99]$\\ %\hline
    
     $100\theta_{\mathrm{MC}}$ & $[0.5, 10]$\\ %\hline
     
     $\tau$ & $[0.01, 0.8]$\\ %\hline

     $\ln (10^{10}A_{\rm s})$ & $[1.61, 3.91]$\\ %\hline

     $n_{\rm s}^{}$ & $[0.8, 1.2]$\\ %\hline

     $N_{\rm eff}$ & $[0.05, 10]$\\ \hline 
    
%    \bottomrule %[2pt]     
    \end{tabular}
%    \end{ruledtabular}
    \label{tab:priors}
\end{table}

In this section, we present the fitting results for the 
four cosmological models by using the latest observational data.\ 
Furthermore, we analyze the effects of different models, and the combination of different datasets on the problem of cosmic tensions.\ 
We will also compare the fitting results by treating  
$N_{\rm eff}^{}$ as a free parameter. There are eight free parameters in our fitting of the HDE and $\Lambda$CDM models where the priors are listed in Table \ref{tab:priors}.

Figure\,\ref{fig:HDENeff_contours} shows the marginalized constraints on cosmological parameters $N_{\mathrm{eff}}$, $c$, $H_0$, $\Omega_{\rm m}$, $\sigma_8$ and $S_8$ of model HDE+$N_{\mathrm{eff}}$. Here the first two parameters are the base parameters, and the $H_{0}$, $\Omega_{\rm m}$, $\sigma_8$ and $S_{8}$ are the derived parameters.
In this analysis, we focus on the effect of the four different dataset combinations.\ We can see from the contours of different colours that, the more data is included, the more precise the constraint becomes.\  Regarding parameter $N_{\mathrm{eff}}^{}$ in Fig.\,\ref{fig:HDENeff_contours}, the green contour ({\it Planck}+BAO) lies on the right of the other contours which contain the ACT data, which suggests that the value of $N_{\mathrm{eff}}^{}$ tends 
to be lower if including the ACT dataset.\ Further information about the constraints on $N_{\mathrm{eff}}^{}$ can be found in Sec.\,\ref{subsec:neff}.The HDE parameter $c$ does not change significantly with these four datasets in Fig.\,\ref{fig:HDENeff_contours}.\ 
For the parameters related to cosmic tensions, $H_0^{}$ and $S_8^{}$, 
the local Hubble constant measurement by Riess \textit{et al.}\,\cite{Riess2022} (R22, blue contours) plays an important role.\  When compared to other datasets like ACT, BAO, and RSD, the inclusion of R22 data leads to a significant increase of $H_0^{}$ and a slight decrease of $S_8^{}$.\ This can be seen from the posterior distribution of $H_0$ and $S_8$ on the far right of Fig.\,\ref{fig:HDENeff_contours} and the corresponding contours, because the blue curve and contours show a notable change compared to other colors.\ 
Regarding $S_8^{}$, the addition of ACT, BAO and RSD had little impact on its constraints.\ In the following, we discuss the resolution of Hubble tension, $S_{8}^{}$ tension, and the constraints of 
$N_{\rm eff}$ respectively.

%{\color{red}{and the comparison with $\Lambda$CDM is shown in Fig.~\ref{fig:Neff}. The HDE parameter $c$ does not change significantly with diverse datasets and its central value varies in the interval [0.5, 0.6]. For the parameters related to cosmic tensions, $H_0$ and $S_8$, the local Hubble constant measurement by Riess \textit{et al.}\cite{Riess2022} i.e. R22 (blue contours) plays an important role. And the difference in the value of $S_{8}=\sigma_{8}\sqrt{\Omega_{\rm m}/0.3}$ does not change significantly with the addition of new data.}}

\subsection{The Hubble Tension}
\label{sec:hubble_tension}
\vspace*{-3mm}

%\textcolor{blue}{}

We first examine the HDE model's capability of reconciling Hubble tension with different dataset combinations.\ To compare it with $\Lambda$CDM model, we plot the posterior distributions in the 
$H_0\!-\!S_8$ plane for the four models $\Lambda$CDM, $\Lambda$CDM+$N_{\mathrm{eff}}$, HDE and HDE+$N_{\mathrm{eff}}$ with two data set combinations 
``\textit{Planck}+ACT+BAO+RSD'' (left panel) and ``\textit{Planck}+ACT+BAO+RSD+R22'' (right panel) 
in Fig.\,\ref{fig:H0_S8}.\ 
In Fig.\,\ref{fig:H0_S8}, the measurements of $S_{8}$ from HSC-Y3 
real-space galaxy correlation function 
($S_{8}\!=\!0.769^{+0.031}_{-0.034}$\,\cite{hsc-y3li}) 
and the R22 result 
($H_{0}\!=\!73.04\pm 1.04\,{\rm km}
{\rm s}^{-1}{\rm Mpc}^{-1}$\,\cite{Riess2022}) 
are shown as the horizontal and vertical gray bands.\ 
The closer the constrained contour is towards the bands, the more reconciliation that the model can do to match the R22 and LSS's $S_{8}$ measurements.\ One can see from both panels of Fig.\,\ref{fig:H0_S8} that, by switching from $\Lambda$CDM model (green) to HDE model (red), the contours move towards the right lower corner, indicating that the HDE model prefers a larger $H_{0}$ value and lower $S_{8}$ value 
(see also Table\,\ref{tab:parameters}).\ Certainly, the reconciliation of HDE does better if R22 data is included (right panel).\ But even without the R22, such trend is still obvious in the left panel.\ 
In both panels, we also include the fitting results of the models including $N_{\mathrm{eff}}^{}$ as a free parameter ($\Lambda$CDM+$N_{\rm eff}$ and HDE+$N_{\rm eff}$), 
showing as the gray and blue contours.\ 
There is no significant effect to relieve the $H_0$ tension, but we notice that HDE+$N_{\mathrm{eff}}$ (blue contours) still provide a better fit than $\Lambda$CDM+$N_{\mathrm{eff}}$ (grey contours).

We plot the marginalized posterior distribution of $H_{0}$ for the four models and two datasets in Fig.\,\ref{fig:H0compare} (D1\,=\,\textit{Planck}+ACT+BAO+RSD, D2\,=\,\textit{Planck}+ACT+ 
BAO+RSD+R22).\ With D1 dataset, $H_{0}$ is measured to be 
$69.70 \pm 1.39\ \mathrm{km\, s^{-1}Mpc^{-1}}$ in the HDE model 
and $67.55 \pm 0.39\,\mathrm{km\, s^{-1} Mpc^{-1}}$ 
in the $\Lambda$CDM model (see also Table\,\ref{tab:parameters}).\ 
Thus, HDE brings the $H_0$ measurement closer to R22, with a tension of 
only $1.92\sigma$ instead of the $4.94\sigma$ tension in the $\Lambda$CDM model\,\footnote{We calculate the distance in the best-fitting value by quantifying the difference of the best-fits with respect to the square root of quadrature sum of each errors, i.e., $(73.04\!-\!69.70)/\sqrt{1.04^2\!+\!1.39^2}\!=\!1.92\sigma$ 
and $(73.04-67.55)/\sqrt{1.04^2\!+\!0.39^2}=4.94\sigma$.}.\ 
By including the R22 in the likelihood, the tension in the HDE model is further reduced.\ The $H_{0}$ is measured to be $71.86 \pm 0.93\,\mathrm{km\,s^{-1} Mpc^{-1}}$, which brings the tension further down to $0.85\sigma$, showing a perfect consistency 
between the CMB measurements and R22.\ 
This comparison is also evident in Fig.\,\ref{fig:H0compare}, 
where $H_{0}$'s posteriors of the HDE (red curves) are always on the right-hand side of $\Lambda$CDM (green curves), and are much closer to the R22 measurement (gray vertical band).

\vspace*{-1.5mm}
\subsection{The $S_8$ Tension}
\vspace*{-1.5mm}
%{\color{red}{YZM: a big question here, in Fig.5 for all ACT results, it seems that you only combines with Planck lensing, so you do not combine Planck CMB at all? Why? This also deviates from the main dataset combination.}}\\

We present in Fig.~\ref{fig:S8constraints} the constraints on $S_8$ in the four models using different weak lensing and CMB datasets, 
where blue horizontal bars are the $\pm 1\sigma$ measurements we take from~\cite{act-dr6}, and red bars are the various constraints by this work.\  We can see that the $S_8^{}$ values are significantly less than $0.8$ by different galaxy lensing observations 
(e.g., HSC-Y3\,\cite{hsc-y3li}, KiDS\,\cite{kids-1000}, and 
DES-Y3\,\cite{des-y3amon}\cite{act-dr6}), but the projected value of the $\Lambda$CDM is $0.831 \pm 0.010$ (\textit{Planck}+ACT+BAO+RSD), which is significantly higher than the lensing measurements.\ If we replace $\Lambda$CDM model by the HDE model, then the $S_{8}^{}$ 
can be brought down to $0.823 \pm 0.011$ (see Table \ref{tab:parameters}).\ Therefore, with respect to the KiDS-1000\,\cite{kids-1000} ($S_8 = 0.759^{+0.024}_{-0.021}$), 
DES-Y3\,\cite{des-y3amon} ($S_8^{}\!=\! 0.759^{+0.025}_{-0.023}$), 
and Hyper Suprime-Cam Year-3 Results (HSC-Y3)\,\cite{hsc-y3li} 
($S_8^{} \!= 0.769^{+0.031}_{-0.034}$), 
the discrepancy can be lowered down to 
$2.56\sigma$, $2.42\sigma$, and $1.57\sigma$, respectively, 
which considerably reduces the $S_{8}^{}$ tension.

\vspace*{1mm}

After further including R22 data, the HDE+$N_{\mathrm{eff}}^{}$ model 
predicts the $S_8^{}$ value as $0.807 \pm 0.011$, 
$N_{\mathrm{eff}}^{}\! =\! 2.89 \pm 0.15$, and the parameter $c$ as $0.52 \pm 0.03$, where the $S_8^{}$ value is even closer to the 
low-level galaxy lensing measurements. Although this change does not entirely resolve the $S_8$ tension, it still indicates that the HDE partially alleviates the $S_8$ tension, and the overall fit is still slightly better than the standard $\Lambda$CDM model.

\subsection{The $N_{\mathrm{eff}}$ Constraints}\label{subsec:neff}
\vspace*{-2mm}

The consideration of $N_{\mathrm{eff}}$ as a varying parameter stems from the substantial deviation between the derived value of $N_{\rm eff}=2.74 \pm 1.3$ \cite{act-dr4} obtained from fitting ACT data with the $\Lambda\mathrm{CDM}$+$N_{\mathrm{eff}}$ model and the theoretical value of 3.046. Our primary aim is to explore whether replacing $\Lambda\mathrm{CDM}$ with HDE can mitigate this disparity, while also assessing its impact on cosmological discrepancies.

We further examine the effect of including $N_{\rm eff}$ as a free parameter in the joint likelihood and the changes in the CMB power spectra.\ What affects the CMB power spectra is the total radiation density:
\begin{eqnarray}
\rho_{\rm r}^{}\,=\,\rho_{\gamma}^{}+\rho_{\nu}^{}\,, 
\label{eq:rho_r}
\end{eqnarray} 
where in natural units,
\beqs 
\begin{eqnarray}
\rho_{\gamma}^{} &=& \frac{\pi^{2}}{15}T^{4}_{\gamma} \,,
\\
\rho_{\nu}^{} &=& \frac{7}{8}\frac{\pi^{2}}{15}N_{\rm eff}^{}
T^{4}_{\nu} \,,
\end{eqnarray}
\eeqs 
are the energy densities of photons and neutrinos.\ 
The factor $7/8$ arises from the Fermi-Dirac distribution, $N_{\mathrm{eff}}^{}$ is the {\it effective} number of neutrino species, which does not need to be an integer~\cite{Komatsu2009,Komatsu2011,Aich2020,Dodelson2020}. In fact, $N_{\mathrm{eff}}=3.046$ in the standard model of particle physics~\cite{Mangano2005,Salas2016,Akita2020,Froustey2020,Bennett2021}.

\vspace*{1mm}

Thus, because neutrinos contribute to the total energy density and change the anisotropic stress of the baryon fluid, to increase $N_{\mathrm{eff}}^{}$ can enhance the expansion rate, reduce the CMB damping tail, and also shift the peak and trough positions of the CMB angular power spectra~\cite{Hou2013,Aich2020}.\ 
Hence, the accurate measurements of CMB power spectra can essentially constrain the $N_{\rm eff}^{}$ value to a high precision, which is a primary science target for this and next generations of CMB experiments\,\cite{plc18-cp,act-dr4,Dutcher2021,Abazajian2019}.

\vspace*{1mm}

In Fig.\,\ref{fig:CMBCell}, we plot the CMB temperature (TT) and E-mode (EE) power spectra in the left and right columns, for both the power spectra measurements and the model predictions, and the relative difference with respect to the fiducial $\Lambda$CDM model. In the upper panel, the gray and red data are the unbinned and binned power spectra respectively. We plot the four models' best-fitting power spectra (Table~\ref{tab:parameters}) with combined ``\textit{Planck}+ACT+BAO+RSD+R22'' dataset and that of $\Lambda$CDM with {\it Planck} 2018 best-fitting value (i.e. {\it Planck} TT+TE+EE+lowE+lensing) in different color and stylish lines. The upper panel shows that all four cosmological models almost overlay on each other, and their fittings to {\it Planck} data are almost indistinguishable. The lower panel of Fig.~\ref{fig:CMBCell} shows the fractional differences of the four models with respect to the {\it Planck} best-fitting $\Lambda$CDM model. One can see that the HDE model (blue solid line) matches the data slightly better than the standard model $\Lambda$CDM (black solid line), because the additional wiggles with the suitable amplitude appeared in the HDE model can better match the up-and-down features of the binned angular power spectra (red data dots) than the $\Lambda$CDM. With the additional $N_{\mathrm{eff}}$, the HDE model is boosted further up (blue dashed line), and the $\Lambda$CDM model is pushed downwards (black dashed line), making the fits less good than that without the $N_{\rm eff}$ parameter.

This situation is also evident in Table~\ref{tab:modelcompare}, where we use the $\chi^2$, Akaike information criterion (AIC) and Bayesian information criterion (BIC) to quantify the ``goodness of fit''. The third column is the minimal $\chi^{2}$ of each model, and $\Delta\chi^2$ (the fourth column) is the relative difference to the best-fitting $\Lambda$CDM model. The total $\chi^2$ of the combined dataset can be written as $\chi^2 = \chi^2_{\it Planck} + \chi^2_{\rm ACT} + \chi^2_{\mathrm{BAO+{RSD}}} + \chi^2_{H_{0}}$. One can see that, by switching from $\Lambda$CDM model to the HDE model, the $\chi^{2}_{\rm min}$ drops by a factor of $-9.352$, which is the result from the better matching of the CMB angular power spectra in Fig.~\ref{fig:CMBCell}. In addition, one can see that adding $N_{\rm eff}$ does not alter the goodness of fit significantly. Because HDE has one additional parameter ($c$ in Eq.~(\ref{eq:rho_de})) than the base $\Lambda$CDM, we also use the AIC and BIC as metrics to compare the models, as they compensate for models with fewer parameters. Here we only need to focus on the relative values, which are defined as $\Delta \mathrm{AIC} = \Delta\chi^2_{\rm min} + 2\Delta k$ and $\Delta \mathrm{BIC} = \Delta\chi^2_{\rm min} + \Delta k\ln N$. Here, $\Delta\chi^2_{\rm min}$ and $\Delta k$ represent the difference in the minimum $\chi^2$ value and the additional number of free parameters compared to the standard model, which is unity in our case. The symbol $N$ denotes the number of data points, for dataset ``{\it Planck}+ACT+BAO+RSD+R22'', $N = 10520$ \footnote{$N = N_{\it Planck} + N_{\rm ACT} + N_{\rm BAO+RSD} + N_{\rm R22}$. For $N_{\it Planck}$, data points combine high-$\ell$ TT ($30\leq \ell \leq 2508$), TE ($30\leq \ell \leq 1996$) and EE ($30\leq \ell \leq 1996$), low-$\ell$ temperature ($0\leq \ell \leq 29$), low-$\ell$ EE ($2\leq \ell \leq 27$) and lensing ($8\leq \ell \leq 400$) spectrum, so $N_{\it Planck} = 2478+1966+1966+29+27+392=6858$; ACT data points from CMB power spectrum on scales of $350 < \ell < 4000$, so $N_{\rm ACT} = 3650$; $N_{\rm BAO+RSD} = 11$ verified as in Table \ref{tab:dataset} and $N_{\rm R22} = 1$. Therefore, $N = 6858+3650+11+1=10520$.}. One can see that, even for AIC and BIC, the HDE model provides a better fit compared to the $\Lambda$CDM model, because the AIC drops by a factor of $-7.352$ and the BIC drops by a factor of $-0.091$. It's worth noting that BIC penalizes model parameters more for larger data volumes compared to AIC. Adding the $N_{\rm eff}$ makes the fit less good because it turns to increase the AIC and BIC values. 

\begin{table}[b]
%    \centering
\caption{Comparison of $\chi^2$ fits for different models.}
\vspace*{3mm}
%    \setlength{\tabcolsep}{8pt}
%    \small
%    \renewcommand{\arraystretch}{1.2}
%    \resizebox{8.6cm}{!}{
    \begin{ruledtabular}
    \begin{tabular}{llcccc}%其中，tabular是表格内容的环境；c表示centering，即文本格式居中；c的个数代表列的个数
%    \toprule[2pt]%设置线宽 
    Date Set & Model & $\chi^2_{\mathrm{min}}$ & $\Delta\chi^2_{\mathrm{min}}$ & $\Delta \mathrm{AIC}$ & $\Delta \mathrm{BIC}$\\ \hline
    \toprule %[2pt]设置线宽
       
    \multirow{4}*{\makecell[l]{\textit{Planck}+ACT\\+BAO+RSD\\+R22}} &  $\Lambda \mathrm{CDM}$ & 3058.77 & 0 & 0 & 0\\ %\hline
    
     ~ &  $\Lambda$CDM+$N_{\mathrm{eff}}^{}$ & 3057.66 & -1.107 & 0.893 & 8.154\\ %\hline
    
     ~ &  HDE & 3049.42 & -9.352 & -7.352 & -0.091\\ %\hline
    
    ~ &  HDE+$N_{\mathrm{eff}}^{}$ & 3049.63 & -9.142 & -5.142 & 0.119\\ %\hline
    
%    \bottomrule %[2pt]     
    \end{tabular}
    \end{ruledtabular}
    \label{tab:modelcompare}
\end{table}

\vspace*{1mm}

We now compare the constraints on $N_{\rm eff}$ value. We compare the fitting results of $N_{\mathrm{eff}}$ in Fig.~\ref{fig:Neff} with different combined datasets and two extension models. The gray vertical dashed line with the
band is the result of constraint from {\it Planck}+ACT data on $\Lambda$CDM+$N_{\rm eff}$ model (ACT DR4~\cite{act-dr4}): $N_{\rm eff}=2.74 \pm 1.3$, where the gray solid line is the prediction of $3.046$ from the standard model of particle physics (SMPH)~\cite{Mangano2005,Salas2016,Akita2020,Froustey2020,Bennett2021}. In Fig.~\ref{fig:Neff}, the change in the $N_{\mathrm{eff}}$ value is insignificant for HDE relative to $\Lambda$CDM (red and grey curves), without taking into account the R22 measurement. Once R22 is added, $N_{\mathrm{eff}}$ is shifted from $2.78 \pm 0.14$ (orange dashed curve) to $3.26 \pm 0.12$ (green dashed curves). But for HDE, $N_{\mathrm{eff}}$ increases from $2.80 \pm 0.16$ (orange solid curve) to $2.89 \pm 0.15$ (green solid curves), which is a much smaller shift than the $\Lambda$CDM. Nonetheless, for all these combined datasets and theoretical models, the resultant constraints on $N_{\rm eff}$ are consistent with SMPH's prediction within $2\sigma$ range.

\vspace*{-1mm}
\section{Conclusions} 
\label{conclusion}
\vspace*{-1mm}

In this work, we revisited the cosmological constraints on the holographic dark energy model and compared its results with the standard $\Lambda$CDM model.\ To constrain the models,
we combined the state-of-the-art cosmological datasets including 
the {\it Planck} 2018 CMB temperature, polarisation and lensing power spectra data with the ACT-DR4 power spectrum, Baryon Acoustic Oscillations (BAO) and Redshift Space Distortion (RSD) from 6dF and SDSS survey, and the Cepheids-SN measurement of $H_{0}$ (R22).\ 
We also treat the effective number of relativistic species ($N_{\mathrm{eff}}$) as a free parameter of fit in the likelihoods to examine whether it alters the fitting results.\ In comparison with the $\Lambda$CDM, we found that the HDE model can relieve the $H_{0}$ and $S_{8}$ tensions to a certain degrees.\ Hence, to respond the three questions raised in Sec.\,\ref{sec:intro}, we conclude as follows:
\begin{itemize}
\item[{\Large $\bullet$}] 
Using the datasets ``\textit{Planck}+BAO+RSD+ACT'', we derived constraints on $H_{0}$ in the $\Lambda$CDM model to be 
$67.55 \pm 0.39\,\mathrm{km\, s^{-1} Mpc^{-1}}$, 
which is $4.94\sigma$ lower than the R22 measurement on $H_{0}$  ($\,=\!73.04\pm 1.04\,{\rm km}\,{\rm s}^{-1} {\rm Mpc}^{-1}$).\ 
With the same dataset, the HDE model brings the value of $H_0$ 
down to $69.70 \pm 1.39\ \mathrm{km\ s^{-1} Mpc^{-1}}$, 
which reduces the $H_0$ tension down to the level of $1.92\sigma$ 
and thus mainly solves the Hubble tension (Figs.\,\ref{fig:H0_S8} and~\ref{fig:H0compare}).
    
After including the R22 dataset in the likelihood, 
the tension in the HDE model is further reduced.\ 
The $H_{0}$ of the HDE model is measured to be 
$71.86 \pm 0.93\,\mathrm{km\,s^{-1} Mpc^{-1}}$, 
which further brings the tension between the HDE prediction and the 
R22 measurement down to $0.85\sigma$, leading to a perfect 
agreement with R22. 

\item[{\Large $\bullet$}] 
By using ``{\it Planck}+ACT+BAO+RSD'' dataset in the HDE+$N_{\mathrm{eff}}$ model, the value of $S_8$ is reduced down to $0.816 \pm 0.012$\,; hence the $S_8^{}$ tension is relieved to 
the $1\sigma$-$2\sigma$ level as compared to the low-redshift gravitational lensing measurement (Fig.\,\ref{fig:S8constraints}).\ This tension is further relieved after including the R22 dataset (Fig.\,\ref{fig:S8constraints}).
    
\item[{\Large $\bullet$}] 
By treating $N_{\mathrm{eff}}^{}$ as a free parameter for fit 
in the likelihood chain, the posterior distribution of 
$N_{\rm eff}^{}$ is around $2.7$-$2.8$, 
depending on the models and the datasets used (Fig.\,\ref{fig:Neff}).\ This constraint is consistent with the standard model value $3.046$ within $2\sigma$ confidence limit, and is on the lower side.

\item[{\Large $\bullet$}] 
Overall, by using the AIC criterion of the goodness-of-fit, we find the HDE model can fit the joint datasets ``{\it Planck}+ACT+BAO+RSD+R22'' to the best extent.
    
%    , HDE and $\Lambda$CDM are comparable. Nevertheless, their results still contain theoretical values in the range of 1-2$\sigma$. In Fig.~\ref{fig:CMBCell} and Table~\ref{tab:parameters}, it is obvious that HDE can match the observed data better than other models. 

\end{itemize}

In the near future, many ongoing surveys such as the Vera C. Rubin Observatory (LSST)\,\cite{LSST2012}, Simons Observatory\,\cite{Ade2019}, and Square Kilometre Array (SKA)\,\cite{SKACosmo2020} 
will provide more fruitful and precise data on galaxies, CMB, 
and large-scale structure.\ We will reexamine the HDE model soon 
with these advanced datasets.

\begin{acknowledgments}
This research is funded by the research program ``New Insights into Astrophysics and Cosmology with Theoretical Models Confronting Observational Data'' of the National Institute for Theoretical and Computational Sciences of South Africa. YZM acknowledges the support from National Research Foundation of South Africa with Grant No. 150580, No. 159044, No. CHN22111069370 and No. ERC23040389081. HJH was supported in part by the National NSFC (under grants 12175136 and 11835005).
\end{acknowledgments}

\appendix
\section*{Results of adding Pantheon+ data}\label{app:pantheon}
    \begin{table*}[t]
%  \centering
\caption{%
Constraints of 
\textit{Planck}+ACT+BAO+RSD+Pantheon+ dataset on the $\Lambda$CDM, $\Lambda$CDM+$N_{\rm eff}^{}$, HDE, and HDE+$N_{\rm eff}^{}$ models.\  The upper block is for the constraints on the fundamental parameters, and the lower block is for the derived parameters.\ The quoted error is given at the $68\%$\, C.L.}
\vspace*{2mm}
\begin{ruledtabular}
\begin{tabular}{lllll}%其中，tabular是表格内容的环境；c表示centering，即文本格式居中；c的个数代表列的个数
    %\toprule[2pt]%设置线宽 
   Parameters & $\Lambda$CDM & $\Lambda$CDM+$N_{\mathrm{eff}}$ & HDE & HDE+$N_{\mathrm{eff}}$\\
\hline     
     %\toprule %[2pt]设置线宽    
%     \midrule %[2pt]
    $\Omega_{\rm b}h^2$ & $0.02235\pm0.00012 $& $0.02212\pm0.00017 $ & $0.02246\pm0.00012 $& $0.02235\pm0.00018 $\\ %\hline 
    
    $\Omega_{\rm c}h^2$ & $0.11986\pm0.00086$ & $0.11542\pm0.00245$ & $0.11801\pm0.00098$& $0.11604\pm0.00241$\\ %\hline    
    
    $100\,\theta_{\rm MC}$  & $1.04108\pm0.00027$ & $1.04160\pm0.00039$& $1.04132\pm0.00027$ & $1.04155\pm0.00037$\\ %\hline   
    
    $\tau$  & $0.0535\pm0.0070$ & $0.0535\pm0.0069$ & $0.0617\pm0.0081$& $0.0610\pm0.0079$\\ %\hline    
    
    $\ln(10^{10}A_{\rm s})$  & $3.050\pm0.014$& $3.037\pm0.016$ & $3.063\pm0.015$ & $3.055\pm0.017$\\ %\hline    
    
    $n_{\rm s}$ & $0.9678\pm0.0034$ & $0.9574\pm0.0065$ & $0.9717\pm0.0036$& $0.9665\pm0.0070$\\ %\hline    
    
    $c$ & - & - & $0.69551\pm0.04075$& $0.68549\pm0.03387$\\ %\hline
    
     $N_{\mathrm{eff}} $ &  $3.046$\,(fixed)  & $2.76\pm 0.15$ & $3.046$\,(fixed) & $2.91\pm 0.15$\\ %\hline    
     \hline
    $H_0[\mathrm{km\,s^{-1}\,Mpc^{-1}}]$ & $67.44 \pm 0.37 $ & $65.70 \pm 1.01 $ & $66.07 \pm 0.77 $ & $65.44 \pm 1.00 $ \\ %\hline   
    
    $\Omega_{\rm de}$ & $0.6858\pm0.0051$ & $0.6798\pm0.0062$ & $0.6767\pm0.0086$ & $0.6752\pm0.0076$ \\ %\hline    
    
    $\Omega_{\rm m}$ & $0.3141\pm0.0051$ & $0.3202\pm0.0062$ & $0.3233\pm0.0086$ & $0.3248\pm0.0076$ \\ %\hline    
    
    $\sigma_8$ & $0.8141\pm0.0057$ & $0.8006\pm0.0092$ & $0.8071\pm0.0082$ & $0.8021\pm0.0098$\\ %\hline    
    
    $S_8\equiv \sigma_{8}\sqrt{\Omega_{\rm m}/0.3}$ & $0.833\pm0.010$ & $0.827\pm0.010$ & $0.837\pm0.011$ & $0.834\pm0.011$\\ %\hline 
    %\hline[2pt]
    %\bottomrule[2pt]     
    \end{tabular}
    \end{ruledtabular}
    \label{tab:parameters_SN}
\end{table*}

\begin{figure}[!htb]
    \centering
    \includegraphics[width=.5\textwidth]{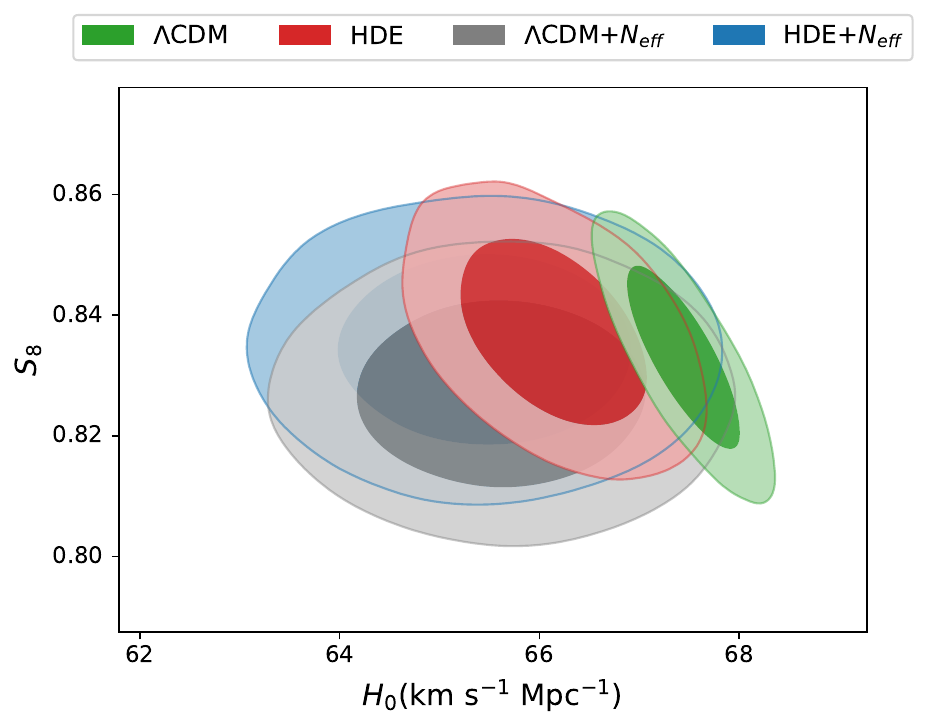}
    \caption{2D marginalized posterior contours in the plane of 
$H_0^{}$ and $S_8^{}$ (68\%\,C.L. and 95\%\,C.L.), 
by using data combination ``\textit{Planck}+ACT+BAO+RSD+Pantheon+'' in the four models $\Lambda$CDM, $\Lambda$CDM+$N_{\mathrm{eff}}$, HDE, and HDE+$N_{\mathrm{eff}}$.}
    \label{fig:H0-S8_SN}
\end{figure}

Here we present the constraints on four models $\Lambda$CDM, HDE, $\Lambda$CDM+$N_{\rm eff}$ and HDE+$N_{\rm eff}$ with dataset ``\textit{Planck}+ACT+BAO+RSD+Pantheon+''. In Fig. \ref{fig:H0-S8_SN}, the posterior contours of $H_0$ and $S_8$ were presented. It is apparent that with the inclusion of the Pantheon+ data, particularly in the $H_0$ comparisons, the models in the HDE scenario do not exhibit a significant advantage over the $\Lambda$CDM. They yield similar results (see Table \ref{tab:parameters_SN}) and slightly larger values of $H_0$ in the $\Lambda$CDM scenario: $H_0 = 67.44 \pm 0.37$ in $\Lambda$CDM compared to $H_0 = 66.07\pm 0.77$ in HDE; and  $H_0 = 65.70 \pm 1.01$ in $\Lambda$CDM+$N_{\rm eff}$ compared to $H_0 = 65.44\pm 1.00$ in HDE+$N_{\rm eff}$. When comparing the constraint on $S_8$ in the $\Lambda$CDM scenario, including Pantheon+ data has little effect on its value. However, in the HDE and extended model, there is an increasing gap with the galaxy lensing observations, as shown in Figs. \ref{fig:H0_S8}(a) and \ref{fig:H0-S8_SN}, and Tables \ref{tab:parameters} and \ref{tab:parameters_SN}, which leads to a similar result in the HDE as in the LCDM with the addition of Pantheon+ data.

Overall, the addition of the Pantheon+ data brings the HDE closer to the $\Lambda$CDM in terms of the constraints on $H_0$ and $S_8$. As discussed in the last part of Section \ref{data} and in Dai et al. (2020)\cite{dai2020}, Pantheon data show a large difference between partial and full redshifts on the constraints on $H_0$ in HDE, with the Pantheon samples having a strong correlation between high and low redshifts. Pantheon+ data likely have a similar effect, and whether there is a unique theoretical reason for this variation in the Pantheon-related data on the constraints of $H_0$ and $S_8$ for the HDE model deserves to be studied and explored in more detail further in the future.

\nocite{*}

\bibliography{HDE}% Produces the bibliography via BibTeX.
\bibliographystyle{apsrev}

\end{document}